\documentclass[letterpaper,twocolumn,english,prl,showpacs]{revtex4}
\usepackage[T1]{fontenc}
\usepackage[latin9]{inputenc}
\usepackage{color}
\usepackage{bm}
\usepackage{amsmath}
\usepackage{amssymb}
\usepackage{graphicx}
\usepackage{esint}
\usepackage{footnote}
\usepackage{CJK}

\makeatletter

\providecommand{\tabularnewline}{\\}

\@ifundefined{textcolor}{}
{%
 \definecolor{BLACK}{gray}{0}
 \definecolor{WHITE}{gray}{1}
 \definecolor{RED}{rgb}{1,0,0}
 \definecolor{GREEN}{rgb}{0,1,0}
 \definecolor{BLUE}{rgb}{0,0,1}
 \definecolor{CYAN}{cmyk}{1,0,0,0}
 \definecolor{MAGENTA}{cmyk}{0,1,0,0}
 \definecolor{YELLOW}{cmyk}{0,0,1,0}
 }

\makeatother

\makeatother

\usepackage{babel}

\begin{document}
\begin{CJK*}{GBK}{song}

\title{Intra-valley Spin-triplet $p+ip$ Superconducting Pairing in Lightly
Doped Graphene}

\author{Jianhui Zhou}

\affiliation{Institute of Physics, Chinese Academy of Sciences, Beijing 100190,
China }

\author{Tao Qin}

\affiliation{Institute of Physics, Chinese Academy of Sciences, Beijing 100190,
China }

\author{Junren Shi}

\affiliation{International Center for Quantum Materials, Peking University, Beijing
100871, China}

\begin{abstract}
We analyze various possible superconducting pairing states and their
relative stabilities in lightly doped graphene. We show that, when
inter-sublattice electron-electron attractive interaction dominates
and Fermi level is close to Dirac points, the system will favor intra-valley
spin-triplet $p+\mathrm{i}p$ pairing state. Based on the novel pairing
state, we further propose a scheme for doing topological quantum computation
in graphene by engineering local strain fields and external magnetic
fields.
\end{abstract}


\maketitle
Graphene has attracted great experimental and theoretical interests
since its successful fabrication.~\cite{nature experiment,BerryEX}
The unique electronic structure of graphene, characterized by massless
relativistic particle-like dispersion near two inequivalent corner
points of Brillouin zone, gives rise to exotic physical properties.~\cite{review}
One of interesting questions would be: how electrons could be paired
in such a system? There had been many theoretical studies that explore
the possibility of the superconductivity in single/multilayer graphenes.
In particular, Ref.~\onlinecite{Neto} predicts a spin singlet $p+\mathrm{i}p$
pairing state in the presence of effective attractive e-e interaction
between the nearest neighboring sites. More exotic possibilities such
as chiral $s$-wave, $d$-wave and $f$-wave pairings are also predicted.~\cite{doniach,Honerkamp,Hosseini,Nandkishore2012,roy1}
First principles calculation has suggested that superconductivity
could be induced by doping alkaline adatoms.~\cite{Profeta2012}
Recently, superconductivity has been realized in potassium-doped few-layer
graphene.~\cite{TBCao}

In this Brief Report, we investigate possibility of obtaining intra-valley
spin-triplet $p+\mathrm{i}p$ superconducting state (see below) in
lightly doped graphene. Unlike the spin-singlet $p+\mathrm{i}p$ pairing
state,~\cite{Neto} the intra-valley spin triplet $p+\mathrm{i}p$
state may support topological excitations of Majorana fermions, and
could be useful for implementing topological quantum computation.
We carry out a systematic investigation on the possible superconductivity
pairing states and their relative stabilities in lightly doped graphene,
taking account of multiple degree of freedoms of orbit, spin and valley.
We find that, when effective $e-e$ attractive interaction between
the nearest neighboring sites (inter-sublattice) dominates and Fermi
level is close to the Dirac points, the intra-valley spin-triplet
$p+\mathrm{i}p$ pairing state will be favored. Based upon the novel
pairing state, we propose a scheme for doing quantum computation in
graphene by engineering local strain fields and external magnetic
fields.

\begin{figure}
\includegraphics[width=1\columnwidth]{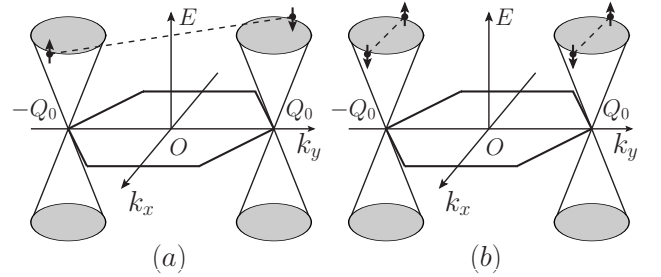} \caption{\label{pairing}Two scenarios of pairings: (a) intervalley pairing,
(b) intravalley pairing. $\bm{Q}_{0}=(0,4\pi/(3\sqrt{3}a))$.}
\end{figure}

We start our investigation from a simplified lattice model introduced
in Ref.~\onlinecite{Neto}, which assumes that effective $e-e$ interaction
has the form:
\begin{multline}
H_{\mathrm{int}}=\frac{g_{0}}{2}\sum_{i,s}\left[a_{is}^{\dagger}a_{is}a_{i-s}^{\dagger}a_{i-s}+b_{is}^{\dagger}b_{is}b_{i-s}^{\dagger}b_{i-s}\right]\\
+g_{1}\underset{<i,j>}{\sum}\underset{ss^{\prime}}{\sum}a_{is}^{\dagger}a_{is}b_{js^{\prime}}^{\dagger}b_{js^{\prime}}\,,\label{eq:Neto}
\end{multline}
where $g_{0}$ and $g_{1}$ denote the strength of onsite interaction
and interaction between the nearest neighboring (NN) sites, respectively.
$a_{is}^{\dagger}$ ($b_{is}^{\dagger})$ and $a_{is}$ ($b_{is})$
are the creation and annihilation operators on the unit cell $i$
with the spin index $s=\uparrow,\downarrow$, and in the sublattice
$A(B)$. $\langle i,j\rangle$ denotes all pairs of NN sites.

We rewrite Eq. (\ref{eq:Neto}) in the basis of non-interacting eigenstates:
$\eta_{l}(\bm{k})=(1/\sqrt{2})\left[1,\,-l\exp\left(-\mathrm{i}\varphi_{\bm{k}}\right)\right]^{T}$,
where $l=1$ ($-1$) for conduction (valance) band, $\exp[i\varphi\left(\bm{k}\right)]\equiv\gamma_{\bm{k}}/|\gamma_{\bm{k}}|$,
$\gamma_{\bm{k}}\equiv\sum_{\bm{\delta}_{i}}e^{i\bm{k}\cdot\bm{\delta}_{i}}$
for $\bm{\delta}_{1}=a\left(3,\sqrt{3}\right)/2$, $\bm{\delta}_{2}=a\left(3,-\sqrt{3}\right)/2$,
$\bm{\delta}_{3}=0$, and $a$ is the C-C bond length. We obtain:
\begin{multline}
H_{\mathrm{int}}=\frac{1}{2N}\sum_{\bm{k}_{i}\bm{K}ss^{\prime}}V^{ss^{\prime}}(\bm{k}_{i})\delta_{\bm{k}_{1}+\bm{k}_{2}-\bm{k}_{3}-\bm{k}_{4}-\bm{K}}\\
\times c_{s}^{\dagger}(\bm{k}_{1})c_{s^{\prime}}^{\dagger}(\bm{k}_{2})c_{s^{\prime}}(\bm{k}_{3})c_{s}(\bm{k}_{4})\,,\label{eq:Hint}
\end{multline}
where $c_{s}^{\dagger}(\bm{k})$ and $c_{s}(\bm{k})$ are creation
and annihilation operators of conduction (or valance) band with wavevector
$\bm{k}$ and spin $s$, respectively, $N$ is the total number of
unit cells in the system, $\bm{K}$ is a reciprocal unit vector, and
the summation of $\bm{k}_{i}$ is over the Brillouin zone. We assume
that inter-band contribution can be ignored, and the band index $l$
is dropped for brevity. The interaction matrix elements have the form:
\begin{multline}
V^{ss^{\prime}}(\mathbf{k}_{i})=\frac{g_{0}}{4}\left[1+e^{\mathrm{i}\varphi_{\bm{k}_{1}}+\mathrm{i}\varphi_{\bm{k}_{2}}-\mathrm{i}\varphi_{\bm{k}_{3}}-\mathrm{i}\varphi_{\bm{k}_{4}}}\right]\delta_{s,-s^{\prime}}\\
+\frac{g_{1}}{4}\left[\gamma_{\bm{k}_{4}-\bm{k}_{1}}e^{\mathrm{i}\varphi_{\bm{k}_{1}}-\mathrm{i}\varphi_{\bm{k}_{4}}}+\gamma_{\bm{k}_{3}-\bm{k}_{2}}e^{\mathrm{i}\varphi_{\bm{k}_{2}}-\mathrm{i}\varphi_{\bm{k}_{3}}}\right]\,.\label{eq:Vss'}
\end{multline}

We focus on the case of lightly doped graphenes with $k_{F}a\ll1$,
where $k_{F}$ is the Fermi momentums of the two Fermi pockets around
the corners of the Brillouin zone at $\pm\bm{Q}_{0}$, as shown in
Fig.~\ref{pairing}. We introduce valley index $\lambda=\pm$ to
denote which valley the electron is in, and the corresponding valley
electron annihilation operator is defined as $c_{\bm{k}\lambda s}\equiv c_{s}(\lambda\bm{Q}_{0}+\bm{k})$,
where $\bm{k}$ is electron momentum relative to the center of the
corresponding valley.

It is easy to see that there could exist two forms of superconducting
pairings in graphene, i.e., inter-valley and intra-valley pairings,
as shown in Fig.~\ref{pairing}. In all the previous studies, only
the inter-valley pairing, i.e., two electrons in a pair have the opposite
momentums and reside at the different valleys (Fig.~\ref{pairing}(a)),
is considered. On the other hand, the intra-valley pairing, i.e.,
two electrons in a pair reside in the same valley, and the two valleys
in the vicinity of $\pm\bm{Q}_{0}$ act like two independent sub-systems
(Fig.~\ref{pairing}(b)), is also possible. To consider both possibilities,
we enumerate all possible combinations of $\bm{k}_{i}$ in the proximity
of $\bm{Q}_{0}$ or $-\bm{Q}_{0}$ in Eq.~(\ref{eq:Vss'}), and collect
all the terms relevant to the superconductivity pairings, i.e., those
terms satisfying either $\bm{k}_{1}+\bm{k}_{2}=\bm{k}_{3}+\bm{k}_{4}=0$
(for the inter-valley pairing) or $\bm{k}_{1}+\bm{k}_{2}=\bm{k}_{3}+\bm{k}_{4}=\pm2\bm{Q}_{0}$
(for the intra-valley pairing). The effective electron-electron interaction
can then be written as,
\begin{multline}
H_{\mathrm{int}}^{\mathrm{sc}}\approx\frac{1}{2N}\sum_{\bm{k}\bm{k}^{\prime}\lambda=\pm,ss^{\prime}}V_{\mathrm{intra}}^{\lambda,ss^{\prime}}(\bm{k},\bm{k}^{\prime})c_{\bm{k}^{\prime}\lambda s}^{\dagger}c_{-\bm{k}^{\prime}\lambda s^{\prime}}^{\dagger}c_{-\bm{k}\lambda s^{\prime}}c_{\bm{k}\lambda s}\\
+\frac{1}{2N}\sum_{\bm{k}\bm{k}^{\prime}\lambda=\pm,ss^{\prime}}\left[V_{\mathrm{inter}}^{ss^{\prime}}(\bm{k},\bm{k}^{\prime})c_{\bm{k}^{\prime}\lambda s}^{\dagger}c_{-\bm{k}^{\prime}\bar{\lambda}s^{\prime}}^{\dagger}c_{-\bm{k}\bar{\lambda}s^{\prime}}c_{\bm{k}\lambda s}\right.\\
\left.+V_{\mathrm{inter}}^{\prime,ss^{\prime}}(\bm{k},\bm{k}^{\prime})c_{\bm{k}^{\prime}\lambda s}^{\dagger}c_{-\bm{k}^{\prime}\bar{\lambda}s^{\prime}}^{\dagger}c_{-\bm{k}\lambda s^{\prime}}c_{\bm{k}\bar{\lambda}s}\right],\label{eq:Hintsc}
\end{multline}
and to the first order of $|\bm{k}|a,|\bm{k}^{\prime}|a$, the matrix
elements can be approximated as:
\begin{align}
V_{\mathrm{intra}}^{\lambda,ss^{\prime}}(\bm{k},\bm{k}^{\prime}) & \approx\frac{g_{0}}{2}\cos(\theta_{\bm{k}}-\theta_{\bm{k}^{\prime}})e^{\mathrm{i}\lambda(\theta_{\bm{k}}-\theta_{\bm{k}^{\prime}})}\delta_{s,-s^{\prime}}\nonumber \\
 & +\frac{3g_{1}}{2}e^{\mathrm{i}\lambda(\theta_{\bm{k}}-\theta_{\bm{k}^{\prime}})}\,,\label{eq:Vintra}\\
V_{\mathrm{inter}}^{ss^{\prime}}(\bm{k},\bm{k}^{\prime}) & \approx\frac{g_{0}}{2}\delta_{s,-s^{\prime}}+\frac{3g_{1}}{2}\cos(\theta_{\bm{k}}-\theta_{\bm{k}^{\prime}})\,,\label{eq:Vinter1}\\
V_{\mathrm{inter}}^{\prime,ss^{\prime}}(\bm{k},\bm{k}^{\prime}) & \approx\frac{g_{0}}{2}\delta_{s,-s^{\prime}}\,,\label{eq:Vinter2}
\end{align}
where we have made use of approximations: $\gamma_{\bm{k}}\approx3$,
$\gamma_{\pm\bm{Q}_{0}+\bm{k}}=\gamma_{\pm2\bm{Q}_{0}+\bm{k}}\approx0$,
$\exp(\mathrm{i}\varphi_{\pm\bm{Q}_{0}+\bm{k}})\approx-\mathrm{i}\exp(\mp\mathrm{i}\mathrm{\theta}_{\bm{k}})$,
to the first order of $ka$, and $\theta_{\bm{k}}$ is the azimuth
of $\bm{k}$.

We can then determine all possible pairing states and their relative
stabilities. First, we consider the case $g_{0}<0$ and $g_{1}>0$.
For the inter-valley pairing, we define the superconducting gap $\Delta_{1}=-(|g_{0}|/2N)\sum_{\bm{k}}\left\langle c_{-\bm{k}-\downarrow}c_{\bm{k}+\uparrow}\right\rangle $,
and $ $$\Delta_{2}=-(|g_{0}|/2N)\,\sum_{\bm{k}}\left\langle c_{-\bm{k}+\downarrow}c_{\bm{k}-\uparrow}\right\rangle $.
It is easy to show that self-consistent gap equations read: $\Delta_{1}=\Delta_{2}=(|g_{0}|/2)\sum_{\bm{k}}\Delta\tanh(\beta E_{\bm{k}}/2)/(2E_{\bm{k}})$,
where $E_{\bm{k}}=\sqrt{(\epsilon_{\bm{k}}-\mu)^{2}+\Delta^{2}}$,
and $\Delta=\Delta_{1}+\Delta_{2}$, $\epsilon_{\bm{k}}=\pm v|\bm{k}|$
is the free electron dispersion of graphene, $\mu$ is the chemical
potential of the system. The equations yield usual $s$-wave pairing
with superconducting gap $\Delta=\Delta_{0}=2\hbar\omega_{0}\exp\left(-1/|g_{0}|N_{0}\right)$
at the zero temperature, where $\hbar\omega_{0}$ is the cut-off energy
of the attractive interaction (e.g., Debye energy in the case of phonon-mediated
interaction) and $N_{0}$ is the density of state per unit cell per
valley at the Fermi level. We note that the pairing correlation function
has even parity when exchanging the valley indexes, and odd parity
when exchanging the spin indexes, respectively, corresponding to a
spin-singlet pairing state $(|+-\rangle+|-+\rangle)(|\uparrow\downarrow\rangle-|\downarrow\uparrow\rangle)$.
On the other hand, there is no pairing in the spin triplet channel.

For the intra-valley pairing, the relevant interaction component is
$U_{\bm{k}\bm{k}^{\prime}}^{\lambda}\equiv-(|g_{0}|/2)\cos(\theta_{\bm{k}}-\theta_{\bm{k}^{\prime}})\exp[\mathrm{i}\lambda(\theta_{\bm{k}}-\theta_{\bm{k}^{\prime}})]$
in Eq.~(\ref{eq:Vintra}). We define the superconducting gap as $\Delta^{\lambda}(\bm{k})=(1/N)\sum_{\bm{k}^{\prime}}U_{\bm{k}\bm{k}^{\prime}}^{\lambda}\left\langle c_{-\bm{k}^{\prime}\lambda\downarrow}c_{\bm{k}^{\prime}\lambda\uparrow}\right\rangle $,
and $\lambda=\pm$. The self-consistent gap equation reads: $\Delta^{\lambda}(\bm{k})=-(1/N)\sum_{\bm{k}^{\prime}}U_{\bm{k}\bm{k}^{\prime}}\Delta^{\lambda}(\bm{k}^{\prime})\tanh(\beta E_{\lambda\bm{k}^{\prime}}/2)/(2E_{\lambda\bm{k}^{\prime}})$
and $E_{\lambda\bm{k}}=\sqrt{(\epsilon_{\bm{k}}-\mu)^{2}+\left|\Delta^{\lambda}(\bm{k})\right|^{2}}$.
The equation has an $s$-wave solution $\Delta^{\lambda}(\bm{k})=\Delta_{0}$
and a $d+\mathrm{i}d$ solution $\Delta^{\lambda}(\bm{k})=\Delta_{0}\exp\left(2\mathrm{i}\lambda\theta_{\bm{k}}\right)$,
with $\Delta_{0}=2\hbar\omega_{0}\exp\left(-4/|g_{0}|N_{0}\right)$.~%
\footnote{The other solutions have the form $\Delta^{\lambda}(\bm{k})=\sqrt{2}\Delta_{0}\cos(\theta_{\bm{k}}+\varphi)\exp\left(\mathrm{i}\lambda\theta_{\bm{k}}\right)$,
and $ $ $\Delta_{0}\approx1.72\hbar\omega_{0}\exp\left(-4/|g_{0}|N_{0}\right)$.
The corresponding states have the higher energy.%
}

Second, we consider the case $g_{0}>0$ and $g_{1}<0$. For the inter-valley
pairing, the relevant interaction component is $U_{\bm{k}\bm{k}^{\prime}}\equiv-(3|g_{1}|/2)\cos(\theta_{\bm{k}}-\theta_{\bm{k}^{\prime}})$
in Eq.~(\ref{eq:Vinter1}). We define the superconducting gap as
$\Delta_{ss^{\prime}}(\bm{k})=-(1/N)\sum_{\bm{k}^{\prime}}U_{\bm{k}\bm{k}^{\prime}}\left\langle c_{-\bm{k}^{\prime}-s^{\prime}}c_{\bm{k}^{\prime}+s}\right\rangle $.
The self-consistent gap equation reads, $\Delta_{ss^{\prime}}(\bm{k})=-(1/N)\sum_{\bm{k}^{\prime}}U_{\bm{k}\bm{k}^{\prime}}\Delta_{ss^{\prime}}(\bm{k}^{\prime})\tanh(\beta E_{\bm{k}^{\prime}}/2)/(2E_{\bm{k}^{\prime}})$
and $E_{\bm{k}}=\sqrt{(\epsilon_{\bm{k}}-\mu)^{2}+\left|\Delta_{ss^{\prime}}(\bm{k})\right|^{2}}$.
The equation yields two stable zero temperature solutions in $p$-wave
channel: $\Delta_{ss^{\prime}}(\bm{k})=\Delta_{0}\exp(\pm\mathrm{i}\theta_{\bm{k}})$
and $\Delta_{0}=2\hbar\omega_{0}\exp[-4/3|g_{1}|N_{0}]$. The pairing
can be either spin singlet or triplet, with the corresponding valley
states $(|+-\rangle-|-+\rangle)$ and $(|+-\rangle+|-+\rangle)$,
respectively.~%
\footnote{The degeneracy between the spin singlet and triplet states will be
lifted in the order of $(k_{F}a)^{2}$, in favor of the singlet state. %
}

For the intra-valley pairing, the relevant interaction component is
$U_{\bm{k}\bm{k}^{\prime}}^{\lambda}=-(3|g_{1}|/2)\exp[\mathrm{i}\lambda(\theta_{\bm{k}}-\theta_{\bm{k}^{\prime}})]$
in Eq.~(\ref{eq:Vintra}). We define the superconducting gap as $\Delta_{ss^{\prime}}^{\lambda}(\bm{k})=(1/N)\sum_{\bm{k}^{\prime}}U_{\bm{k}\bm{k}^{\prime}}^{\lambda}\left\langle c_{-\bm{k}^{\prime}\lambda s^{\prime}}c_{\bm{k}^{\prime}\lambda s}\right\rangle $.
The self-consistent gap equation reads $\Delta_{ss^{\prime}}^{\lambda}(\bm{k})=-(1/N)\sum_{\bm{k}^{\prime}}U_{\bm{k}\bm{k}^{\prime}}^{\lambda}\Delta_{ss^{\prime}}^{\lambda}(\bm{k}^{\prime})\tanh(\beta E_{\bm{k}^{\prime}}/2)/(2E_{\bm{k}^{\prime}})$
and $E_{\bm{k}}=\sqrt{(\epsilon_{\bm{k}}-\mu)^{2}+\left|\Delta_{ss^{\prime}}^{\lambda}(\bm{k})\right|^{2}}$.
The equation yields the spin triplet $p+\mathrm{i}p$ pairing state
at zero-temperature: $\Delta_{ss^{\prime}}^{\lambda}=\Delta_{0}\exp(\mathrm{i}\lambda\theta_{\bm{k}})$
and $\Delta_{0}=2\hbar\omega_{0}\exp(-2/3|g_{1}|N_{0})$.

The above considerations assume that the Fermi level is close enough
to the Dirac points, such that the quasi-particle dispersion has perfect
rotation symmetry within each valley. However, real graphene systems
only have three-fold rotation symmetry. The asymmetry, i.e., $\epsilon_{\pm\bm{Q}_{0}+\bm{k}}\ne\epsilon_{\pm\bm{Q}_{0}-\bm{k}}$,
becomes more pronounced when Fermi level is removed from the Dirac
points, leading to pair breaking that destabilizes the intra-valley
pairing. On the other hand, the inter-valley pairing is not affected
because the symmetry between $\epsilon_{\bm{k}}$ and $\epsilon_{-\bm{k}}$
is guaranteed by time-reversal symmetry. To see asymmetry effect to
the intravalley pairings, we expand the tight-binding dispersion to
the second order, and for the valley centering $\bm{Q}_{0}$, we have
$\epsilon_{\bm{k}}^{v}\equiv\epsilon_{\bm{Q}_{0}+\bm{k}}\approx\hbar vk+h_{\bm{k}}$,
and $h_{\bm{k}}=(\hbar k^{2}va/4)\sin3\theta_{\bm{k}}$. The gap equation
for the intra-valley pairing when $g_{1}<0$ is modified to $\Delta_{ss^{\prime}}^{\lambda}(\bm{k})=-(1/N)\sum_{\bm{k}^{\prime}}U_{\bm{k}\bm{k}^{\prime}}^{\lambda}\Delta_{ss^{\prime}}^{\lambda}(\bm{k}^{\prime})[1-f(E_{\bm{k}^{\prime}}+h_{\bm{k}^{\prime}})-f(E_{\bm{k}^{\prime}}+h_{-\bm{k}^{\prime}})]/(2E_{\bm{k}^{\prime}})$.
Solving the equation, we find that the intravalley pairing states
become unstable when $\mu>2\sqrt{(\hbar v/a)\Delta}\approx130\sqrt{\Delta/1\mathrm{meV}}\,\mathrm{meV}$
for graphene.

The discussion can be further extended to the more general form of
the interaction. We introduce $g_{\alpha\beta}^{ij}$ as e-e interaction
between two electrons resided in the unit cell $i$ sublattice $\alpha$
and the unit cell $j$ sublattice $\beta$ with $\alpha,\:\beta\in\left\{ A,\: B\right\} $.
The Fourier transformations are $\tilde{g}_{0}(\bm{q})=(1/N)\underset{ij}{\sum}g_{\alpha=\beta}^{ij}\exp\left[-i\bm{q}\cdot(\mathbf{R}_{i}-\mathbf{R}_{j})\right]$
and $\tilde{g}_{1}(\bm{q})=(1/N)\underset{ij}{\sum}g_{\alpha\neq\beta}^{ij}\exp\left[-i\bm{q}\cdot(\mathbf{R}_{i}-\mathbf{R}_{j})\right]$,
corresponding to intra-sublattice and inter-sublattice interactions,
respectively. In the limit of $k_{F}a\ll1$, only Fourier components
at $\bm{q}=0$ and $\bm{q}=\pm2\bm{Q}_{0}$ are relevant. Moreover,
we can show that $\tilde{g}_{1}(\pm2\mathbf{Q}_{0})$ always vanishes
because of $C_{3}$ rotational symmetry. The interaction matrix elements
in Eq.~(\ref{eq:Hintsc}) are then modified to:
\begin{align}
V_{\mathrm{intra}}^{\lambda,ss^{\prime}}(\bm{k},\bm{k}^{\prime}) & \approx\frac{\tilde{g}_{0}(\bm{0})}{2}\cos(\theta_{\bm{k}}-\theta_{\bm{k}^{\prime}})e^{\mathrm{i}\lambda(\theta_{\bm{k}}-\theta_{\bm{k}^{\prime}})}\nonumber \\
 & +\frac{\tilde{g}_{1}(\bm{0})}{2}e^{\mathrm{i}\lambda(\theta_{\bm{k}}-\theta_{\bm{k}^{\prime}})}\,,\label{eq:Vintra-1}\\
V_{\mathrm{inter}}^{ss^{\prime}}(\bm{k},\bm{k}^{\prime}) & \approx\frac{\tilde{g}_{0}(\bm{0})}{2}+\frac{\tilde{g}_{1}(\bm{0})}{2}\cos(\theta_{\bm{k}}-\theta_{\bm{k}^{\prime}})\,,\\
V_{\mathrm{inter}}^{\prime,ss^{\prime}}(\bm{k},\bm{k}^{\prime}) & \approx\frac{\tilde{g}_{0}(2\bm{Q}_{0})}{2}\,.\label{eq:V-1}
\end{align}

Following the same procedure, we can extend our analysis to the general
model. The results are summarized in Table~\ref{tab:Pairing states}.
For the simple model Eq.~(\ref{eq:Neto}), $\tilde{g}_{0}(\bm{0})=\tilde{g}_{0}(2\bm{Q}_{0})=g_{0}$,
and $\tilde{g}_{1}(0)=3g_{1}$. Compared to the simple model, the
general model yields extra possibility of forming inter-valley spin
triplet $s$-wave pairing state.

\begin{table}
\begin{tabular}{|c|c|c|c|c|}
\hline
 & \multicolumn{2}{c|}{Intervalley} & \multicolumn{2}{c|}{Intravalley ($\mu<2\sqrt{\frac{\hbar v\Delta}{a}}$)}\tabularnewline
\hline
\hline
$\triangle(\bm{k})$  & $1$ & $\begin{array}{c}
\exp(\mathrm{i}\theta_{\bm{k}})\\
\exp(-\mathrm{i}\theta_{\bm{k}})
\end{array}$ & $\begin{array}{c}
1\\
\exp(2\mathrm{i}\lambda\theta_{\bm{k}})
\end{array}$ & $\exp(\mathrm{i}\lambda\theta_{\bm{k}}$)\tabularnewline
\hline
valley & $\pm$  & $\pm$ & $+$ & $+$\tabularnewline
\hline
spin  & $\mp$ & $\pm$ & $-$ & $+$\tabularnewline
\hline
$V_{0}$ & $\frac{\left|\tilde{g}_{0}(\bm{0})\pm\tilde{g}_{0}(2\mathbf{Q}_{0})\right|}{2}$  & $\frac{\left|\tilde{g}_{1}(\bm{0})\right|}{4}$  & $\frac{\left|\tilde{g}_{0}(\bm{0})\right|}{4}$  & $\frac{\left|\tilde{g}_{1}(\bm{0})\right|}{2}$ \tabularnewline
\hline
\end{tabular}

\caption{\label{tab:Pairing states}The possible superconducting pairing states.
The momentum dependence of gap function $\Delta(\bm{k})$, the parity
in exchanging the valley ({}``valley'') and spin ({}``spin'')
index, as well as the effective pairing strength $V_{0}$ are shown.
The amplitude of the zero temperature superconducting gap is related
to $V_{0}$ by $\Delta_{0}=2\hbar\omega_{0}\exp(-1/N_{0}V_{0})$. }
\end{table}

\begin{figure}
\includegraphics[width=1\columnwidth]{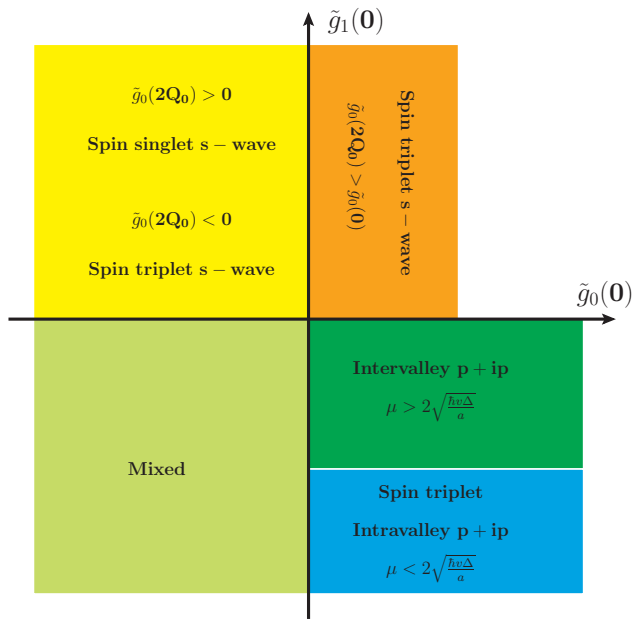}

\caption{\label{fig:phase diagram}Phase diagram of lightly doped graphene.
For given $\tilde{g}_{0}(\bm{0})$ and $\tilde{g}_{1}(\bm{0})$, the
pairing states also depend on the parameters $\tilde{g}_{0}(2\bm{Q}_{0})$
and $\mu$, and the corresponding conditions are shown for each phase. }
\end{figure}

From Table \ref{tab:Pairing states}, it is straightforward to determine
the pairing ground states in different parameter regimes. The phase
diagram is shown in Fig.~\ref{fig:phase diagram}. The most interesting
feature of the phase diagram is the presence of spin-triplet intra-valley
$p+\mathrm{i}p$ pairing state when the inter-sublattice attractive
interaction $\tilde{g}_{1}(\bm{0})$ dominates, and the Fermi level
is close to the Dirac points such that $\mu\leq2\sqrt{(\hbar v/a)\Delta}$.
It is important to note that the spin-triplet intra-valley $p+\mathrm{i}p$
pairing state does not break the time reversal symmetry (TRS), because
the order parameters at the two valleys are related by the time reversal,
i.e., the order parameter at one valley is $p+\mathrm{i}p$, while
that at the other valley is $p-\mathrm{i}p$. The origin of the intra-valley
chiral $p+\mathrm{i}p$ pairing can be traced back to $\pi$-Berry
phase associating with the Dirac point in each of the valleys.~\cite{Shi2006,Mao2011}
The topological origin of the unconventional pairing, which is dictated
by a band singularity buried deeply inside the Fermi sea, is different
from usual mechanisms derived from strong correlations. \textcolor{black}{The
intra-valley chiral $p+\mathrm{i}p$ superconducting state is spatially
non-uniform, due to the finite }center of mass momentum (CMM) of Cooper
pair\textcolor{black}{s. It is important to note that the pairing
with the finite CMM does not incur extra kinetic energy in graphene,
because the electron dispersion does have minimums at $\pm\bm{Q}_{0}$
(see Fig.~\ref{pairing}). This is distinct from the another well
known non-uniform superconducting state in conventional systems, i.e.,
the LOFF state,~\cite{LO,FF} which incurs large kinetic energy and
is very fragile.}

The intra-valley triplet $p+\mathrm{i}p$ superconducting phase in
lightly doped graphene, once realized, could be used to implement
graphene-based topological quantum computation. This could be achieved
if one could generate half quantum vortices (HQV) in the superconducting
phase.~\cite{Jang2011} Each vortex core will confine a Majorana
fermion. The quantum state of a macroscopic system with a number of
such vortices could be transformed by adiabatically moving them, with
robustness topologically protected by their spatial separations.~\cite{Ivanov2001}

For graphene, there exists a very useful and unique tool for implementing
the novel quantum computing scheme, i.e., pseudo-magnetic fields generated
by strains or topological defects.~\cite{Guinea2009,Voz2010,Levy2010}
Using a strain field or topological defect, it is possible to create
a spatially localized pseudo-magnetic flux, which is not easily achievable
using a real magnetic field. One can use the pseudo-magnetic fluxes
as nuclei for creating HQVs. For instance, one can generate a total
pseudo-magnetic flux of a half quanta within the size of superconducting
penetrating length. Because of the presence of two independent valleys,
it will create a pair of HQVs, one for each of the valleys. The two
Majorana fermions localized inside the vortex cores will maintain
their independence within time scale of inter-valley scattering.~\cite{Wu2007}
On the other hand, to create a single Majorana fermion, one could
engineer a total pseudo-magnetic flux of a quarter quanta. Because
the pseudo-magnetic field has the opposite directions for the two
valleys, a perpendicular real external magnetic field will enhance
the magnetic field experienced by one valley while suppressing it
in the other. It is then possible to generate a HQV only in one of
the two valleys.~%
\footnote{To minimize elastic energy, it is usually necessary to create pseudo-magnetic
fluxes in pairs of opposite directions. In this case, each pair of
the fluxes will create a pair of Majorana fermions residing in the
different valleys, spatially separated. %
} The resulting Majorana fermion will be fully topologically protected.
Finally, one could carefully design the strain field and the matching
external real magnetic field so that the creation of multiple HQVs
is energetically favorable,~\cite{Chung2007} and the graphene-based
topological quantum computation can then be achieved by tuning the
strain field.

To realize the novel superconducting phase predicted in this Report,
it is important to find a modification to graphene systems that introduces
an appropriate level of carrier density and a sizable effective electron-electron
attractive interaction from such as electron-phonon coupling or electron-plasmon
coupling.~\cite{Neto} Recently, it has been predicted by the first
principles calculation that graphene could be transformed superconducting
by lithium deposition.~\cite{Profeta2012} While the particular system
has its electronic structure significantly modified, and cannot be
expected to support the novel $p+\mathrm{i}p$ phase we predict, it
does give the hope that one may find other dopants that serve the
purpose.

In summary, we systematically investigate the superconducting phases
for lightly doped graphene. We find that an intra-valley spin triplet
$p+\mathrm{i}p$ superconducting phase could be achieved in  the certain
parameter regime. We further show that the novel superconducting phase
could be used to implement topological quantum computation in graphene
by utilizing pseudo-magnetic fluxes created by strain field or topological
defects. These possibilities make graphene a promising candidate for
implementing topological quantum computation.

This work is supported by MOST 973 program No.~ 2009CB929101.

\end{CJK*}
\end{document}